# Electronic bandstructure of in-plane ferroelectric van der Waals β´-In$_2$Se$_3$

James L. Collins[1,2,3*], Chutian Wang[3,4], Anton Tadich[3,5], Yuefeng Yin[3,4], Changxi Zheng[1,2,3], Jack Hellerstedt[1,2,3], Antonija Grubišić-Čabo[1,2,3], Shujie Tang[6], Sung-Kwan Mo[6], John Riley[7], Eric Huwald[7], Nikhil V. Medhekar[3,4], Michael S. Fuhrer[1,2,3] & Mark T. Edmonds[1,2,3*]

[1]School of Physics and Astronomy, Monash University, Clayton, Victoria, Australia.

[2]Monash Centre for Atomically Thin Materials, Monash University, Clayton, Victoria, Australia.

[3]ARC Centre for Future Low Energy Electronics Technologies, Monash University, Clayton, Victoria, Australia.

[4]Department of Materials Science and Engineering, Monash University, Clayton, Victoria, Australia

[5]Australian Synchrotron, Clayton, Victoria, Australia.

[6]Advanced Light Source, Lawrence Berkeley National Laboratory, Berkeley, CA, USA.

[7]Department of Physics, La Trobe University, Bundoora, Victoria, Australia

*email: james.collins@monash.edu and mark.edmonds@monash.edu

*Abstract -*

Layered indium selenides (In$_2$Se$_3$) have recently been discovered to host robust out-of-plane and in-plane ferroelectricity in the α and β´ phases, respectively. In this work, we utilise angle-resolved photoelectron spectroscopy to directly measure the electronic bandstructure of β´-In$_2$Se$_3$, and compare to hybrid density functional theory (DFT) calculations. In agreement with DFT, we find the band structure is highly two-dimensional, with negligible dispersion along the c-axis. Due to *n*-type doping we are able to observe the conduction band minima, and directly measure the minimum indirect (0.97 eV) and direct (1.46 eV) bandgaps. We find the Fermi surface in the conduction band is characterized by anisotropic electron pockets with sharp in-plane dispersion about the $\overline{M}$ points, yielding effective masses of 0.21 m$_0$ along $\overline{KM}$ and 0.33 m$_0$ along $\overline{\Gamma M}$. The measured band structure is well supported by hybrid density functional theory calculations. The highly two-dimensional (2D) bandstructure with moderate bandgap and small effective mass suggest that β´-In$_2$Se$_3$ is a potentially useful van der Waals semiconductor. This together with its ferroelectricity makes it a viable material for high-mobility ferroelectric-photovoltaic devices, with applications in non-volatile memory switching and renewable energy technologies.

**INTRODUCTION**

Layered van der Waals materials such as III−VI binary chalcogenides (MX or $M_2X_3$, with M = Ga or In, and X = S, Se or Te) possess a diverse range of crystalline phases and structures leading to a wide range of electronic and optical properties that hold great promise for technological application in electronics and optoelectronics [1-9]. Of particular interest is the recent discovery that the α- and β´-phases of $In_2Se_3$ host out-of-plane [10] and in-plane [11] ferroelectricity. This ferroelectric behaviour, coupled with high photo-responsivity [12] and a bandgap in the infrared/visible range makes $In_2Se_3$ ideal for two-dimensional (2D) ferroelectric applications [5,13,14]. Importantly, fundamental electronic properties in these phases of $In_2Se_3$ such as the carrier effective mass, as well as direct and indirect electronic bandgaps have yet to be reliably measured experimentally. This, along with wide-ranging values reported for the optical bandgap between 1.2 eV and 1.8 eV [15-18], mandates a precise and direct experimental determination of the electronic structure of $In_2Se_3$.

In this work, we exploit angle-resolved photoelectron spectroscopy (ARPES) to directly measure the occupied electronic bandstructure to determine both the direct and indirect bandgaps and the carrier effective masses. The ARPES results on bulk crystals of β´-$In_2Se_3$ reveals the valence band maximum is slightly offset from the $\bar{\Gamma}$ point, whilst the conduction band minima resides at the $\bar{M}$ point yielding an indirect bandgap of 0.97 eV and a direct bandgap of 1.46 eV. The valence band curvature is relatively flat, corresponding to a large hole effective mass, whilst, the sharply dispersing conduction band yields a small electron effective mass of 0.21 $m_0$ along $\overline{KM}$ and 0.33 $m_0$ along $\overline{\Gamma M}$ directions due to the band anisotropy. Density function theory calculations performed on the undistorted hexagonal unit cell β-$In_2Se_3$ show excellent agreement with these observations, suggesting that there is minimal disturbance in the electronic structure of β´-$In_2Se_3$ with respect to the β-phase.

Importantly, the electron effective mass we measure in β´-In$_2$Se$_3$ is lower than the values reported for transition-metal dichalcogenides such as MoS$_2$ (0.4-0.6 m$_0$) [19], is slightly larger than blue phosphorene (m$_e^*$~0.13 m$_0$) [20], black phosphorus (m$_x^*$=0.15 m$_0$) [21] and InSe (0.14-0.16 m$_0$) [1]. The above materials possess room-temperature (RT) electron mobilities of $10^3$-$10^4$ cm$^2$V$^{-1}$s$^{-1}$ [1,22], suggesting achieving ultra-high mobility in In$_2$Se$_3$ is also possible, and is crucial in developing high efficiency opto-electronic and ferroelectric devices based on β´-In$_2$Se$_3$.

**RESULTS & DISCUSSION**

The unit cell of β-In$_2$Se$_3$ is made up of planar quintuple layers (QLs) with strong in-plane covalent bonding and weak out-of-plane van der Waals interactions. Figure 1(a) shows a schematic of a 3-quintuple layer unit cell structure of β-phase In$_2$Se$_3$, where each QL consists of Se−In−Se−In−Se planes with a total thickness of ~1 nm. Around the c-axis there is three-fold rotational symmetry. Figure 1(b) shows a schematic of the bulk and surface-projected Brillouin zones (BZ) of β-phase In$_2$Se$_3$. The observed in-plane ferroelectricity manifests due to the formation of one-dimensional (1D) periodic modulations along these high-symmetry directions; we adopt the nomenclature of Ref. 11 and refer to this distorted ferroelectric phase as the β´-phase, which has a reduced symmetry that more closely approximates a hexagonal space group. We note that a recent publication on 2D In$_2$Se$_3$, uses the label β´-phase to refer to a completely new crystal structure.[8]

We first measure the atomic and crystal structure of in-situ cleaved β´-phase In$_2$Se$_3$ using scanning tunnelling microscopy (STM) and low-energy electron diffraction (LEED) in Fig. 1(c) and (d) respectively. The STM topograph in Fig. 1(c) reveals the 3-5 dimer-row distortions attributed to the

1D-striped ferroelectric phase [11], and a triangular arrangement of selenium atoms (marked as red circles for clarity). Local fluctuations in the apparent topographic height may be the result of charge inhomogeneity, which is not uncommon in polar surfaces. The LEED results in Fig. 1(d) confirm the 1×1 structure due to the absence of superstructure spots, whilst energy-dependent LEED (not shown) reveals the expected three-fold rotational crystal symmetry. The absence of additional LEED spots corresponding to the three distinctly oriented ferroelectric domains as previously observed in LEEM [11], indicates the domains are averaged over due to the several hundred micron LEED spot size.

We first use ARPES to confirm that the electronic structure shows the same periodicity and symmetry as observed in LEED. We probe the reciprocal-space distribution of electronic states at a fixed binding energy [$E_F$ - 1.6 eV corresponding to the valence band maximum (VBM)] taken at $hv = 100$ eV at room temperature (RT). The results are shown in Fig. 1(e). The unique toroidal analyser geometry [23] enables detection of the full photoemission hemisphere, which coupled with the higher photon energy samples a wide region in k-space encompassing multiple Brillouin zones. In Fig. 1(e) the hexagonal surface BZ boundaries are overlaid in white. The constant energy surface shows the same clear three-fold rotational symmetry as the LEED, confirming that the electronic structure has to a good approximation the symmetry of the undistorted hexagonal β-phase $In_2Se_3$, and the long-wavelength (small wavevector) modulations of the electronic structure due to the ferroelectric distortion are minimal. Therefore, from now on we will present our experimental results using the surface Brillouin zone and high-symmetry points of the undistorted β-phase. We also compare our measurements directly to first-principles DFT calculations of the undistorted hexagonal β-phase $In_2Se_3$ crystal structure without accounting for electronic distortions due to the ferroelectricity in the β´-phase.

We now turn to measurements of the band dispersion below Fermi level (E$_F$) along the $\overline{\Gamma M}$, $\overline{MK}$ and $\overline{\Gamma K}$ high-symmetry directions at 110K. Figure 2(a) shows energy distribution curves (EDCs) along these high-symmetry directions (i.e. binding energy as a function of wavevector parallel to the surface, k$_\parallel$) taken with *p*-polarized incident light at *hv* = 80 eV. It was found that *p*-polarized light was required to highlight the conduction band state, as the spectral intensity of the conduction band was only barely observable using *s*-polarized incident light. In our experimental setup s- and p-polarization correspond to the electric field vector E of the incident radiation perpendicular and parallel to the detection plane of the analyser; this difference strongly affects the optical dipole selection rules affecting emission intensity [24]. The experimental data in Fig. 2(a) is overlaid with DFT calculated band structures using a Heyd-Scuseria-Ernzerhof (HSE) hybrid functional where the red and turquoise solid lines reflect $k_z$ = 0 and $k_z = \pm\pi/c$ respectively. The HSE indirect bandgap is 0.64 eV (k$_z$=0); in order to match the experimental data to compare to the band dispersion in Fig. 2(a) the HSE bandgap was increased by 0.33 eV. The close similarity of the calculated bands at $k_z$ = 0 and $k_z = \pm\pi/c$ indicates that the bandstructure shows little dispersion along $k_z$, i.e. strongly two-dimensional with less dispersion in k$_z$ than other 2D van der Waals semiconductors such as MoS$_2$ [25]. In principle our experimental bandstructure is measured at a particular $k_z$ which is a function of photon energy as well as in-plane momentum and could be estimated within e.g. the free-electron final-state approximation. Nevertheless, as discussed below, any dispersion along $k_z$ is below our experimental resolution and given our conclusions are insensitive to the precise determination of $k_z$, except for an uncertainty in band positions of order ±0.1 eV.

Bands dispersing downwards (consistent with the conduction band) are present at $E_F$ albeit at a much weaker intensity that the valence band features, thus independent colour-scaling (normalization) of the conduction and valence bands was performed on each (green boxed) section of the composite plot to properly highlight the conduction band on the same figure as the valence band. The observation of electron pockets at $E_F$ indicates the In$_2$Se$_3$ crystals are heavily *n*-type doped, consistent with earlier findings [26] and are likely the result of bulk vacancies arising from the crystal growth process [16], and may also be due to the large number of surface defects observed in large area STM topography (not shown). A faint electron pocket centred at $\bar{\Gamma}$ is evident and extends 0.15 eV below $E_F$, whilst a strong sharply dispersing band at $\bar{M}$ extends 0.77 eV below $E_F$ representing the conduction band minimum (CBM). Figure 2(b) shows energy distribution curves (i.e. intensity as a function of binding energy) extracted at the $\bar{\Gamma}$ (blue) and $\bar{M}$ (red) points which were used to help determine the CBM. The mostly flat valence band displays a 'Mexican hat' structure, where the valence band maximum bulge resides ~0.2 Å$^{-1}$ from the zone centre 1.76 eV below $E_F$. From the positions of the CBM and VBM (red and light-blue dot-dashed lines respectively in Fig. 2(a)) β´-In$_2$Se$_3$ is an indirect band gap semiconductor. The transition energy for the indirect gap is $\Delta_{\Gamma-M} = 1.74 - 0.77 = 0.99 \pm 0.10 \; eV$, whilst the lowest direct gap occurs at the $\bar{M}$ point with $\Delta_{M-M} = 2.23 - 0.77 = 1.53 \pm 0.10 \; eV$ and finally a direct gap at the $\bar{\Gamma}$ point of $\Delta_{\Gamma-\Gamma} = 1.92 - 0.15 = 1.77 \; eV \pm 0.10 \; eV$. Values for the experimental and theoretical bandgaps are summarized in Table 1.

We now turn to photon-energy dependent ARPES to probe whether β´-In$_2$Se$_3$ exhibits any appreciable $k_z$-dependence, in order to confirm its 2D or 3D electronic nature. Figure 2(c) shows momentum distribution curves taken at the Fermi energy for photon energies ranging between 56.5 eV and 90 eV. The position and size of the electron pocket at $k_{||} = 0$ ($\bar{\Gamma}$, normal emission) does not change with photon energy, and simply shows a decrease in intensity as the photon energy is increased, which is consistent with the quasi-two-dimensional behaviour expected for van der Waals layered materials. This minimal $k_z$ dispersion is also consistent with the DFT band structure calculations for the $k_z = \pm\pi/c$ and $k_z = 0$ planes shown as red and turquoise lines in Fig. 2(a) respectively.

Overall, we see excellent agreement between experiment and calculations with DFT reproducing the key experimental features such as the electron pockets at the $\overline{M}$ point, the higher energy $\overline{\Gamma}$ point electron band and the overall valence band structure. However, the energy and momentum broadening in our experiment prevent direct measurement of the DFT predicted sub-band splitting around the $\overline{M}$ point.

We now turn to high-resolution measurements of the constant energy contours of the conduction and valence band. Figure 3 shows constant energy maps taken using ARPES ($hv$ = 80 eV) at (a) $E_F$ and (b) $E_F$ - 1.76 eV that are overlaid with the DFT calculated isoenergies (grey overlay represents all $k_z$), with the Brillouin zone marked by black lines. In Fig. 3(a) an anisotropic Fermi surface with near three-fold rotational symmetry is observed that is preserved for all energies down to the CBM and is in excellent agreement with DFT. The slight variations of photoemission intensity may be attributed to momentum-sensitive optical selection rules. The observation that β´-In$_2$Se$_3$ retains its three-fold symmetry at RT (see Fig. 1(e)) and at 110 K (see Fig. 3(a) and also STM taken at 77 K see Fig. 1(c)) indicates the bulk crystals do not undergo the same temperature-dependent phase transformation as observed recently in 2D In$_2$Se$_3$ [8], or that the transition temperature is significantly lower in bulk crystals. Figure 3(b) shows a constant energy map taken at $E_F$ – 1.76 eV corresponding to the VBM. The most obvious features are a clear decrease in the intensity of the bands at $\overline{\Gamma}$ and a bulb-like extrusion structure around it, with peaks in the intensity occurring at 0.2 Å$^{-1}$ along the $\overline{\Gamma M}$ direction, again in excellent agreement with DFT.

We now examine the detailed dispersion of the conduction band, to determine the effective mass, which will govern the charge carrier mobility in β´-In$_2$Se$_3$. Figure 4 shows high-resolution ARPES spectra along (a) $\overline{\Gamma M}$ and (b) $\overline{KM}$, where white profiles reflect momentum distribution curves taken at $E_F$ and the purple triangles correspond to peak positions extracted from Gaussian fits from both momentum and energy distribution curves. Figure 4(c) and (d) compare the peak positions extracted from experiment with the DFT calculated band dispersions for $k_z = 0$ along the $\overline{\Gamma M}$ and $\overline{KM}$ directions respectively. A considerable broadening of the bands is evident (white profile), that exceeds the experimental energy and momentum resolution broadening, and is most likely due to a combination of crystal surface quality, charge disorder and lifetime broadening coupled with the small but finite $k_z$ dispersion. Despite this, the experiment and theory show reasonable agreement in the overall band curvature in particular along the $\overline{KM}$ direction. In order to calculate the electron effective mass ($m_e^*$) we fit the conduction bands along the two high symmetry directions to a nearly free electron model (solid purple lines in Fig. 4(c) and (d)) given by $E = \frac{\hbar^2 k^2}{2m_e^*}$, to within a 0.3 Å$^{-1}$ region extending from the $\overline{M}$ point. Figure 4(e) re-plots the experimental data as $E$ vs. $k^2$, along $\overline{\Gamma M}$ (red) and $\overline{KM}$ (turquoise) directions. This clearly demonstrates a linear relationship, i.e. a parabolic dispersion and constant effective mass, extending to momenta well away from the high symmetry point. Considering the uncertainty of each peak position due to the energy/momentum broadening, we obtain experimental $m_e^*$ values of $0.328 \pm 0.010 m_0$ for $\overline{\Gamma M}$ and $0.208 \pm 0.01 m_0$ for $\overline{KM}$ respectively. These values are in reasonable agreement with the DFT predicted effective masses of $0.39 \pm 0.05 m_0$ ($\overline{\Gamma M}$) and $0.27 \pm 0.05 m_0$ ($\overline{KM}$), and the agreement could be further improved with $k \cdot p$ perturbation theory calculations to map the band dispersion close to the high symmetry points. It should also be noted that an effective mass of 0.24 $m_0$ was inferred from electrical measurements on In$_2$Se$_3$ single crystals [26], which further supports our findings of low carrier effective mass. Values for the experimental and theoretical effective masses are summarized in Table 1.

**CONCLUSION**

In summary we have used ARPES to clarify the electronic structure of in-plane ferroelectric β´-In$_2$Se$_3$ to show that it is an indirect bandgap semiconductor. The indirect $\overline{\Gamma M}_{VBM} \rightarrow \overline{M}_{CBM}$ 0.97 eV transition and $\overline{M}_{VBM} \rightarrow \overline{M}_{CBM}$ 1.46eV direct transition make this material suitable for optoelectronic applications in the near-infrared region. The nearly 2D bandstructure observed experimentally and predicted theoretically suggests that the bandgap will increase only modestly when β´-In$_2$Se$_3$ is reduced in thickness by exfoliation [8] or growth [27] of atomically thin In$_2$Se$_3$. Additionally, we found that the low-energy carriers are typified by a surprisingly low effective mass, ranging between 0.21 to 0.33m$_0$. The retention of large bandgap and small effective mass for *n*-type carriers is therefore promising for achieving high mobility, suggesting applications in high mobility ferroelectric photovoltaics and non-volatile switching technologies could be realized with $\beta'$-In$_2$Se$_3$.

## METHODS

**Experimental -**Bulk single-crystal samples of high-purity (99.995%) 2H phase β´-In$_2$Se$_3$ (HQ-Graphene) were cleaved in ultra-high vacuum, with structural characterisation undertaken using scanning tunnelling microscopy (STM) at 77 K and low-energy electron diffraction (LEED) at room-temperature. ARPES measurements sampling a large *k*-space range were measured using the Toroidal angle resolving endstation at the Soft X-Ray beamline of the Australian Synchrotron at a photon energy of 100 eV using linearly polarized light at normal incidence to the sample at room temperature. High-resolution ARPES measurements between 40 eV and 90 eV with s- and p- polarized light was performed at Beamline 10.0.1 of the Advanced Light Source (ALS) using a Scienta R4000 analyser at 110 K to avoid charging due to carrier freeze out. All data shown from the ALS was measured using horizontally polarised light. The binding energy scale for all spectra are referenced to the Fermi energy ($E_F$), determined using the Fermi edge of Au foil reference samples. The total convolved energy resolution for measurements taken on the Toroidal Analyser was 100 meV, and for measurements on Beamline 10.0.1 the convolved energy and angular resolution was 35 meV and 0.2º (i.e. 0.016 Å$^{-1}$ for photoelectrons at 80 eV).

**First Principles Density Functional Theory Simulations –** All first principles density functional theory simulations were conducted within the generalised gradient approximation of the form as implemented in Vienna Ab-Initio Simulation Package (VASP) [28,29,30,31]. The plane wave energy cut off was set to 400eV and a dense 21×21×3 Gamma-cantered k-point grid was used to sample the BZ of bulk $In_2Se_3$ unit cell for an accurate description of the electronic structure. The crystal structure was fully optimised until Hellman-Feynman ionic forces were less than 0.01 eV/Å. Finally, the band structures thus obtained were further confirmed with those obtained using high accuracy HSE06 hybrid functionals [32].

**Acknowledgements** M.T.E. was supported by ARC DECRA fellowship DE160101157. J.L.C., M.S.F., Y. Y and N. V. M. acknowledge funding support from Australian Research Council Centre for Excellence Future Low Energy Electronics Technologies (CE170100039). J.L.C. and M.S.F. are supported by M.S.F.'s ARC Laureate Fellowship (FL120100038). M.T.E. and A.T. acknowledge travel funding provided by the International Synchrotron Access Program (ISAP) managed by the Australian Synchrotron, part of ANSTO, and funded by the Australian Government. Part of this research was undertaken on the soft X-ray beamline at the Australian Synchrotron, part of ANSTO. This research used resources of the Advanced Light Source, which is a DOE Office of Science User Facility under contract no. DE-AC02-05CH11231. Y.Y. And N.M. Gratefully acknowledge the computational support from Australian National Computing Infrastructure and Pawsey Supercomputing Facility.

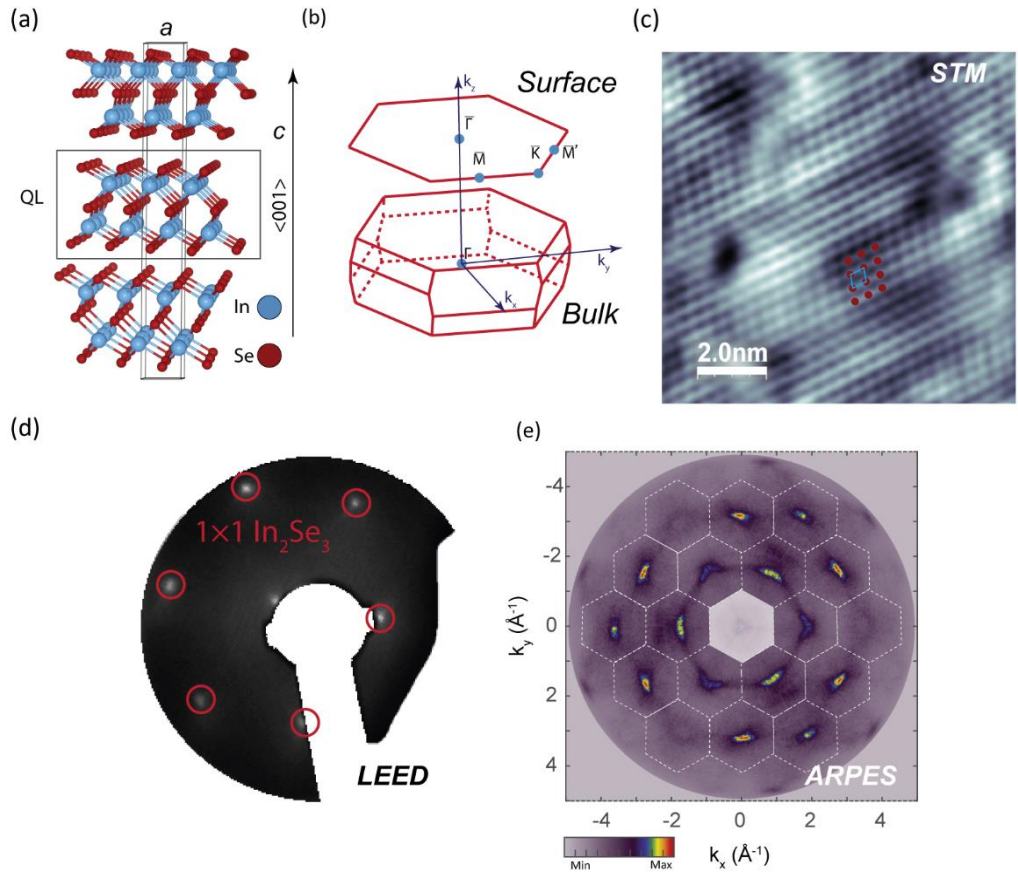

**Fig. 1 | $\beta' - In_2Se_3$ crystal and electronic structure characterization.** (**a**) Crystal structure of $\beta - In_2Se_3$, with indium atoms and selenium atoms in blue and red, respectively. Each quintuple layer (QL) contains five atomic layers in the order of Se–In–Se–In–Se layers. Three QLs form a unit cell. (**b**) Bulk and <001> surface projected Brillouin Zone (BZ) schematic with surface high symmetry points labelled. Labels for bulk points are not included. (c) STM constant-current topographic image ($V$ = 1.7 V, $I$ = 70 pA, $T$ = 77 K) of in-situ cleaved (001) $\beta' - In_2Se_3$ surface showing one-dimensional ferroelectric lattice distortions. Blue rhombus indicates the lattice $1 \times 1$ unit cell (Se atoms marked as red spots as guide to the eye.) (d) Low-energy electron diffraction (LEED) pattern of the cleaved (001) surface at room temperature; red circles mark the 1 × 1 lattice. (**e**) Constant energy in plane momentum map of the photoemission intensity at $E = E_F - 1.6\ eV$ with $hv$ = 100 eV. BZ boundaries are delineated by the white hexagonal grid overlay.

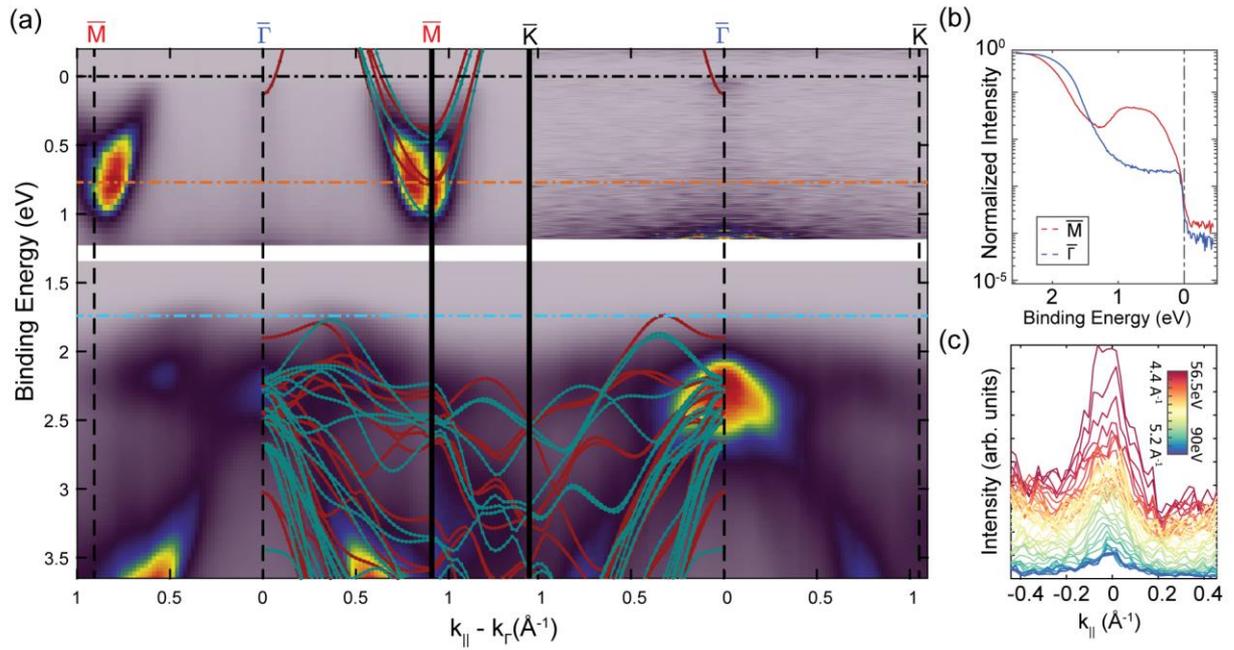

**Fig. 2 | Electronic structure of $\beta' - In_2Se_3$.** (a) Energy distribution curves showing the band dispersion and energy gap along the labelled high-symmetry directions measured at $hv$ = 80 eV. DFT band structures of undistorted hexagonal β-In₂Se₃ for $k_z = 0$ and $k_z = \pm\frac{\pi}{c}$ are overlaid in red and turquoise respectively. The conduction band minima are represented by the orange dashed line and the valence band maxima are represented by the blue dashed line. Each boxed section of (a) has been independently colour scaled in order to highlight the weak intensity conduction band features. (b) Photoemission intensity profiles as function of binding energy at $\bar{\Gamma}$ and $\bar{M}$. (c) Momentum distribution curves at the Fermi-energy taken between $hv$ = 56.5 - 90 eV. Colourscale inset indicates the $k_z$ correspondence for $hv = 56.5 - 90\ eV$ energy range.

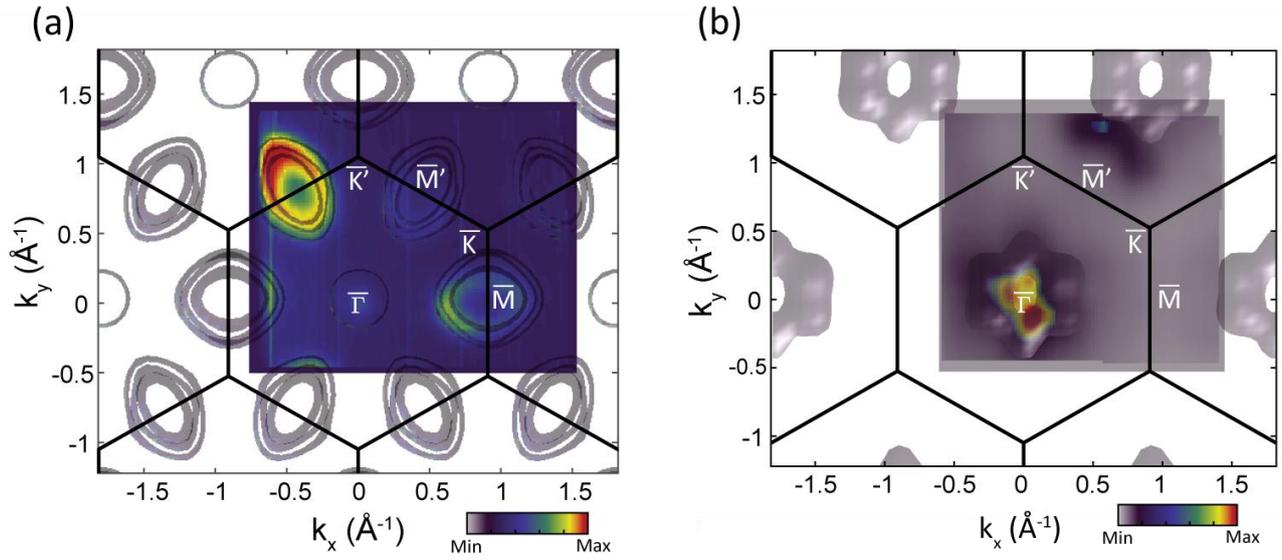

**Fig. 3 | Constant energy mapping of electron pocket Fermi surface and valence band maxima of $\beta' - In_2Se_3$.** (a) Fermi surface map taken at 80 eV highlighting three-fold symmetric geometry and anisotropic band warping. Constant iso-energy DFT solutions are overlaid in grey for all $k_z$. High-symmetry point markers are indicated by the white text overlay. (b) Constant energy map measured at $E_F - 1.76\ eV$ shows the momentum-resolved Fermi surface near the valence band maximum, which is seen to peak away from $\bar{\Gamma}$.

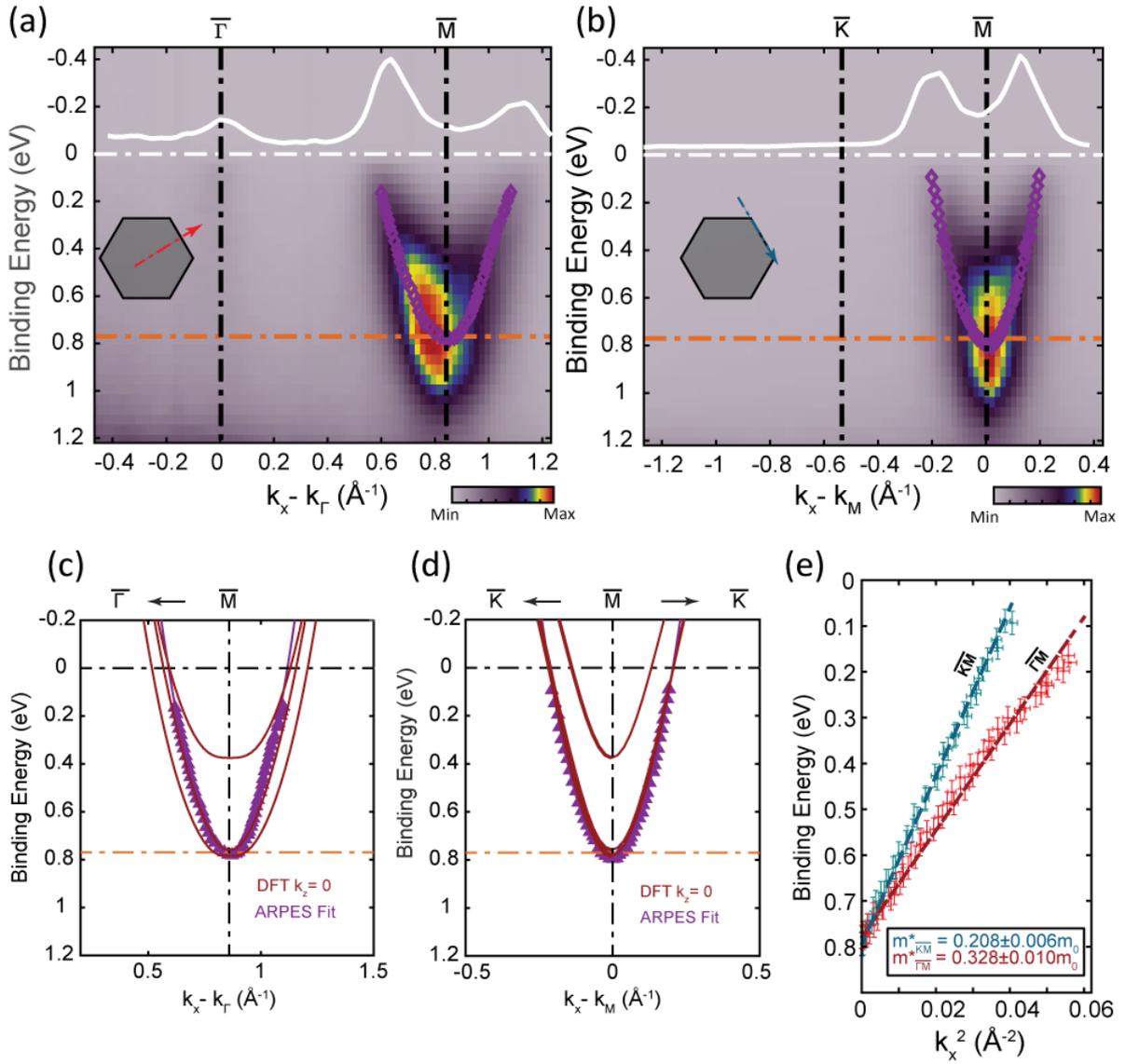

**Fig. 4 | Conduction band effective mass of $\beta' - In_2Se_3$.** Photoemission intensity plots showing energy distribution curves along the (a) $\overline{\Gamma M}$ and (b) $\overline{KM}$ high-symmetry directions. Solid white trace profiles photoemission intensity to highlight conduction pockets at $\overline{\Gamma}$ and around $\overline{M}$ at $E_F$. Purple diamonds mark band locations extracted by analysis of EDCs. The orange dashed line marks the conduction band minimum. Scan direction indicated by inset. (c) and (d) shows the band locations extracted from (a) and (b) as purple diamonds along with the DFT bands (red lines) for $k_z = 0$ overlaid for direct comparison. (e) E vs k² plot of the experimental data in (c) and (d) along $\overline{\Gamma M}$ and $\overline{KM}$ and the extracted effective mass along the high symmetry directions.

Table 1. Bandgap and effective mass values of $\beta' - In_2Se_3$.

|  | $E_{Indirect}$ (eV) | $E_{Direct\ at\ M}$ (eV) | $E_{Direct\ at\ \Gamma}$ (eV) | $m_e^*$ ($\overline{MK}$) | $m_e^*$ ($\overline{\Gamma M}$) |
|---|---|---|---|---|---|
| Experiment | $0.97 \pm 0.1$ eV | $1.46 \pm 0.1$ eV | $1.77 \pm 0.1$ eV | $0.21 \pm 0.01 m_0$ | $0.33 \pm 0.01 m_0$ |
| Theory ($k_z$=0) | 0.64 eV | 1.17 eV | 0.93 eV | $0.27 \pm 0.10 m_0$ | $0.39 \pm 0.10 m_0$ |

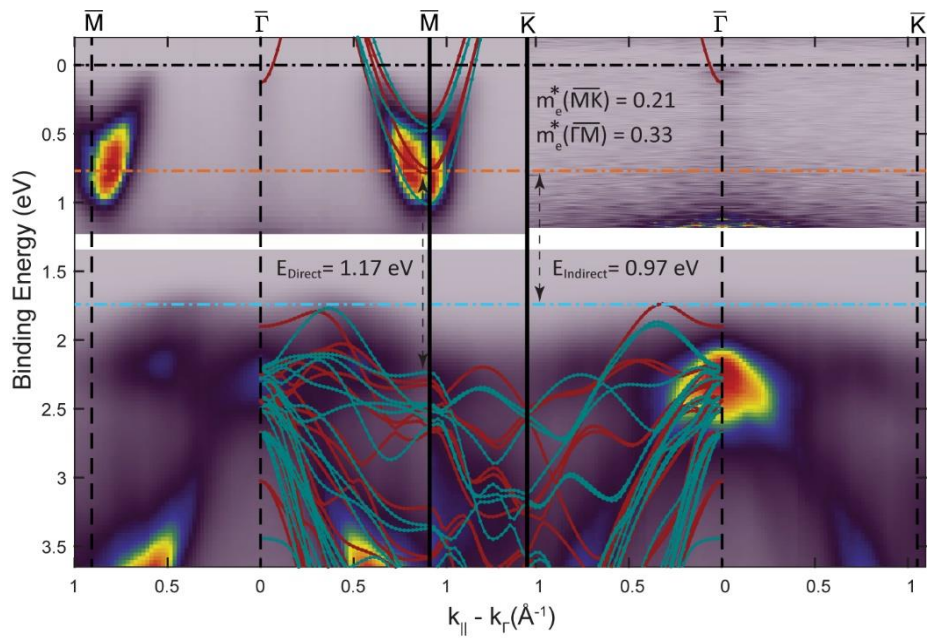

**TOC Graphic**